\documentclass{article}
\usepackage{smc}
\usepackage{times}
\usepackage{ifpdf}
\usepackage[english]{babel}
\usepackage{cite}
\usepackage{enumitem}
\setitemize{noitemsep,topsep=0pt,parsep=0pt,partopsep=0pt}

\def\papertitle{FOR LISA. A PIANO-BASED SONIFICATION PROJECT OF GRAVITATIONAL WAVES}
\def\firstauthor{Andrea Valle}
\def\secondauthor{Valeriya Korol}

\newif\ifpdf
\ifx\pdfoutput\relax
\else
   \ifcase\pdfoutput
      \pdffalse
   \else
      \pdftrue
\fi

\ifpdf 
  \usepackage[pdftex,
    pdftitle={\papertitle},
    pdfauthor={\firstauthor, \secondauthor},
    bookmarksnumbered, 
    pdfstartview=XYZ 
   ]{hyperref}

  \usepackage[pdftex]{graphicx}
  \graphicspath{{./figures/}}
  \DeclareGraphicsExtensions{.pdf,.jpeg,.png}

  \usepackage[figure,table]{hypcap}

\else 
  \usepackage[dvips,
    bookmarksnumbered, 
    pdfstartview=XYZ 
  ]{hyperref}  

  \usepackage[dvips]{epsfig,graphicx}
  \graphicspath{{./figures/}}
  \DeclareGraphicsExtensions{.eps}

  \usepackage[figure,table]{hypcap}
\fi

\hypersetup{
    colorlinks,%
    citecolor=black,%
    filecolor=black,%
    linkcolor=black,%
    urlcolor=black
}

\title{\papertitle}

\twoauthors
   {\firstauthor} {CIRMA/StudiUm - Università di Torino \\ %
     {\tt \href{mailto:andrea.valle@unito.it}{andrea.valle@unito.it}}}
   {\secondauthor} {Institute for Gravitational Wave Astronomy\\ 
   \& School of Physics and Astronomy, 
   \\University of Birmingham \\
     {\tt \href{mailto:korol@star.sr.bham.ac.uk}{korol@star.sr.bham.ac.uk}}}

\begin{document}
\capstartfalse
\maketitle
\capstarttrue
\begin{abstract}
In the paper we discuss the sonification of the simulated gravitational wave data for the future LISA space mission. First, we introduce the LISA project and its output. Then, we present {\em Einstein's Sonata}, a multimedia project devoted to the artistic public display of the LISA data.  {\em Einstein's Sonata} features as its main element a music composition for prepared piano, {\em Periplo del latte}. The latter results from a sonification strategy mapping astronomical data onto music. We thus detail a four-stage sonification procedure, that features two data preprocessing stages, a mapping into an abstract control space and finally automatic notation generation.
\end{abstract}

\section{Introduction}\label{sec:introduction}
Recently astronomers started exploring the Universe through a new kind of radiation: gravitational wave radiation or gravitational waves. Gravitational waves represent perturbations of the space-time generated by masses in acceleration. Their properties are similar to those of electromagnetic waves: they are traversal waves that propagate at the speed of light and can be characterized by properties like frequency and amplitude. Unlike electromagnetic waves that can be seen as visible light or felt as heat, gravitational waves have been conceived theoretically and cannot be perceived directly. Thus, sonification of gravitational waves provides a way of experiencing and interpreting these otherwise imperceptible waves.

In this work we consider gravitational waves detectable with the future {\it Laser Interferometer Space Antenna} (LISA), an European Space Agency-led mission.
Consisting of three identical spacecrafts in an equilateral triangle configuration distant 2.5\,Mkm apart and connected by laser links, LISA will be sensitive to a part of gravitational wave spectrum between $10^{-4} - 10^{-1}$\, Hz ~\cite{LISA}. A large variety of astrophysical sources emit gravitational waves in this frequency band ranging from stellar remnants (black holes, white dwarf and neutron stars) in binary systems in our Milky Way to nascent massive black holes early in the history of the Universe. However, gravitational wave signals from binaries composed of two white dwarf stars -- hereafter double white dwarf (DWD) -- will outnumber all the other astrophysical gravitational wave sources in the LISA frequency band by at least an order of magnitude, and, thus, are the focus of our work. As many as tens of millions of DWDs are expected to emit gravitational waves the LISA band, out of which tens of thousand will be detected and characterized by LISA~\cite{Korol2017}.\\
The estimated number of LISA detectable DWDs is based on theoretical studies of these binaries in the Milky Way realized in preparation for the LISA mission by combining many numerical techniques. The first is to model the intrinsic properties of DWDs -- binary orbital geometry and white dwarf masses that determine the strengths of its gravitational wave signal -- with a technique called binary population synthesis \cite{seba}. It represents a collection of numerical prescriptions (typically motivated by available electromagnetic observations) for processes involved in stellar and binary evolution. These prescriptions are combined in a numerical code that ``evolves'' a binary system from the birth of two stars until they turn into white dwarfs \cite{Nelemans2001,Toonen2012}. Second, a detailed Milky Way model describing its shape is used to realise a mock galaxy \cite{Korol2019}. Practically, it serves to assign 3D positions to synthetic DWDs with respect to LISA that also affects the strength of the gravitational wave signal, which is inversely proportional to the DWD distance from the LISA detector. Next, knowing the position and properties of DWDs, their gravitational wave signals can be fully modelled and the detectability of binary can be assessed by processing them with the LISA detection pipeline, which simulates LISA's response to the incoming gravitational wave \cite{Littenberg2011}.
Lastly, detectable DWDs are collected in a catalogue listing properties for each binary (Section \ref{sec:data}). A similar catalogue can be expected as part of data products generated by the LISA mission. 

As communication of scientific results to a general audience is becoming increasingly important, we developed the multimedia project, {\em Einstein's Sonata},  with the aim of providing an artistic rendition that could express in a creative and accessible way (a part of) the output of the LISA mission.  
{\em Einstein’s Sonata} was born from the idea of transforming theoretical models of gravitational waves signals emitted by DWDs populating our Milky Way into musical and visual arts. Based on the concept of music being a powerful tool of engagement, the goal of this project is to create music by transforming astronomical data into sounds and to contribute to the cultural heritage of the community focusing on both sciences and arts. The visual part of {\em Einstein’s Sonata}, conceived by the artist Samantha Stella, is inspired by the opening scene of the film {\em Drowning by Numbers} (1988) by British film director and artist Peter Greenaway. The film opens with a little girl - adorned in a dress from Spanish Baroque painter Diego Velazquez’s {\em Las Meninas}, 1656 – rope-skipping while counting stars from 1 to 100. The performance of the {\em Sonata} opens with Samantha Stella reading the names of the 100 brightest stars in the Milky Way symbolising future ``brightest'' gravitational waves sources that will be discovered by LISA.
The performance acts as a framework around the key element, a piece for prepared piano entitled {\em Periplo del latte} composed following a sonification strategy that takes into account the experimental data.
The title of the project, {\em Einstein's Sonata}, represents a homage to Albert Einstein, who postulated the existence of gravitational waves in the framework of the theory of General Relativity dated 1916 \cite{Einstein1916}. However, the (direct) detection of gravitational waves became possible almost 100 years later, when Laser Interferometer Gravitational Wave Observatory (LIGO), a ground-based gravitational wave detector, caught a signal generated by the pair of colliding black holes nearly 1.3 billion light years away \cite{ligoGW150914}. 

In the following, we describe the sonification strategy~\cite{HermannHuntNeuhoff2011-SHB} that was used to compose {\em Periplo del latte}, the piece for prepared piano. Again, the title (``periplus of the milk'') is a reference in Italian to the exploration of the space of the Milky Way.

\section{Data and format} \label{sec:data}

Gravitational wave signals emitted by DWDs can be considered as quasi-monochromatic and can be fully described by 8 parameters: amplitude, frequency, frequency derivative (change of the signal's frequency with time), sky coordinates in ecliptic coordinate system (i.e. a spherical coordinate system centred on the position of the Sun and aligned with the ecliptic), binary orbital inclination, gravitational wave polarisation and initial orbital phase. The detectability of each DWD signal is assessed based on the signal-to-noise ratio (SNR), which expresses the strength of the gravitational wave signal with respect to the LISA instrumental noise and can be interpreted as a measure of relevance. In the context of the LISA mission, SNR $=7$ is typically required to detect a monochromatic source, while a higher threshold guarantees the measurement of all eight parameters describing the signal in addition to the detection. The chosen threshold yields a catalogue of $\sim26$k DWDs.

\begin{table}
\centering
\label{tab:my-table}
\begin{tabular}{ll}
\hline
Quantity                         & Units         \\ \hline
frequency $f$                        & Hz            \\
frequency derivative $\dot{f}$             & Hz/s          \\
amplitude  $\cal{A}$                      & dimensionless \\
chirp mass  ${\cal M}$                     & M$_\odot$     \\
distance  $d$                       & kpc           \\
sky coordinates $(\theta, \phi)$ & radiants      \\
signal-to-noise ratio SNR                              & dimensionless \\ \hline
\end{tabular}
\caption{A table of DWD parameters listed in the input data file used for sonification.}
\end{table}

For sonification proposes we consider LISA detectable DWDs (i.e. those with SNR $>7$) only and use parameters listed in Table \ref{tab:my-table}.
Note that this list ignores some of the astrophysically less interesting parameters and explicitly specifies binary's chirp mass -- a combination of the binary components masses that drives frequency evolution of the signal -- and distance that can be derived from the measurement of the amplitude and frequency derivative \cite{Korol2021}.

To summarize:
\begin{itemize}
    \item the input data do not represent a time series, rather a set of measurements (in a model) in relation to points in a space;
    \item in terms of dimension, the catalog -- even if already resulting from a selection based on SNR -- contains more than $26$k measurements, each provided with its $8$ parameters.
\end{itemize}

\begin{figure}[t]
\centering
\includegraphics[width=1\columnwidth]{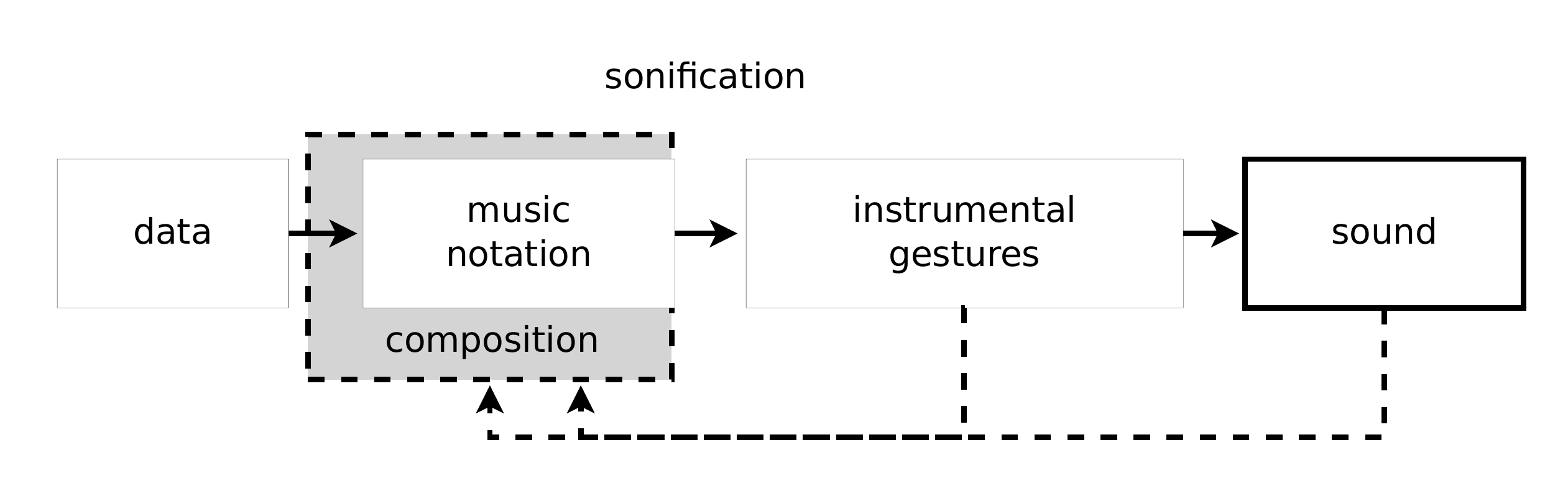}
\caption{Sonification and music composition.\label{fig:soncomp}}
\end{figure}
While there were no strict constraints on the commission side, it was evident that a methodologically-wise design was required in order to respect the scientific nature of the project while composing the music. Thus, we defined a data sonification design. The sonification of astronomical data has a long tradition \cite{VariableStarsMusic,AudioUniverse,Bardelli2021}. An up-do-date discussion with references to previous cases is provided by ~\cite{Bardelli2021} in the context of their sonification of zCOSMOS, an astronomical dataset that contains information about $\sim20$k galaxies. In all these projects the typical output is MIDI or electronic sound, as it is easier to define various mapping strategies within a completely digital environment. Differently from these examples, we decided to compose for piano, which constituted the main constraint of our project, so to be able to propose the result in an acoustic concert setting and as a part of the physical setting of {\em Einstein's Sonata}.

Instrumental music composition can be thought of as a special case of sound design to be realized by mapping sonic features to instructions for musicians, the latter acting as acoustic sources. In this framework, music can be thought of as a sonification of gestures~\cite{DelleMonache2008}. Following this, algorithmic composition can be considered even more strictly a form of sonification, as in this case abstract procedures and data are formally mapped into instrumental gestures. On the other hand, instrumental gestures cannot be addressed directly, rather they require the mediation of music notation. Thus, in the case of music composition for acoustic instruments, a sonification procedure must result in graphic symbols, rather than ending with audio signals. To sum up, far from being immediate, sonification-based instrumental composition implies a chain of transformation processes (Figure \ref{fig:soncomp}).
A composition work (dashed lines) that is exclusively oriented towards acoustic instruments properly deals with the process involving the first two domains (data and music notation), the second being its output. Instrumental gestures and sound, and the process linking them, have to be taken into account as possible constraints while composing, in a feedback loop. \\

\section{A four-stage sonification scheme}
\begin{figure}[t]
\centering
\includegraphics[width=1\columnwidth]{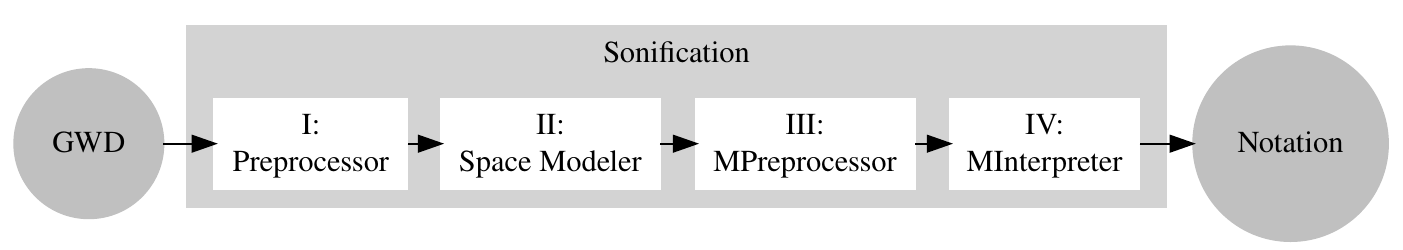}
\caption{Four-stage sonification architecture.\label{fig:general}}
\end{figure}

In relation to the previously discussed sonification chain, we thus designed a pipeline linking data to music notation (Figure \ref{fig:general}). In this parameter-mapping sonification~\cite{Hermann:2000}, gravitational wave data (GWD) transit through a four-stage sonification architecture so to end into music notation. Notation acts as a further symbolic, graphical, layer yet to be sonified by the piano player, finally resulting into sound. The four stages can be grouped into two subsets. Stages I and II map GWD into an abstract space model, while stages III and IV take the output of III in order to automatically generate music notation. Each subset features as its first stage a Preprocessor that filters and maps data so that they can be used in the subsequent one. The ratio for splitting the {\em data $\rightarrow$ notation} process into two subsets lies in the data themselves: as discussed, they do not represent a time series, rather a spatial configuration. Thus, in stages I and II we first construct an abstract space that can be explored. Then, after obtaining a set of time series from such explorations, we map them into music notation (stages III and IV). In short, stages I and II are related to ``space modeling'' while III and IV deal with ``music interpretation'' (see Figures \ref{fig:detailSpace} and \ref{fig:detailInt}, discussed later). The final result is a piece made up of 11 short sections representing into notation different explorations of the data space.  

\section{Space modeling}
\begin{figure}[t]
\centering
\includegraphics[width=1\columnwidth]{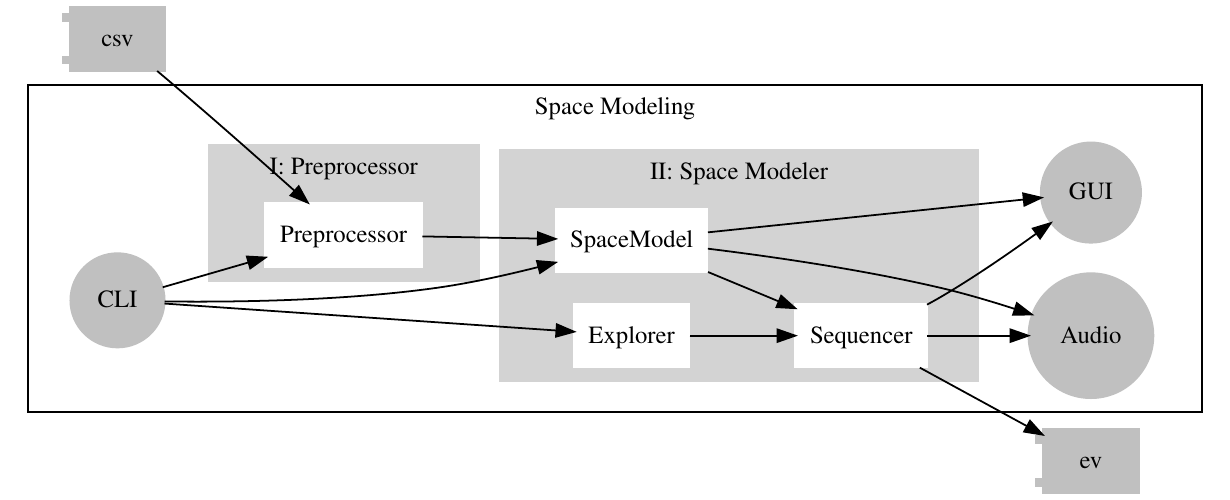}
\caption{Stages I and II: Space modeling.\label{fig:detailSpace}}
\end{figure}

Stages I and II deal with space modeling. The aim of space modeling is to obtain a simplified 2D model that can be explored interactively so to gain information to feed composition. Space modeling (stages I and II) is entirely implemented in the SuperCollider interpreted language~\cite{Wilson:2011}, that provides facilities both for data manipulation and audio and visual displaying.
A detailed description is given in Figure \ref{fig:detailSpace}, to which we refer in the following.

\subsection{Stage I: Preprocessing}
GWD are passed to the Preprocessor via a file in CSV format.
The main issue in the sonification design is the large amount of data to be taken care of. This is particularly relevant if we consider that the final output is music notation, to be passed to the piano player. The graphic output cannot be simply a graphical data visualization, rather it is constrained by a maximum complexity in relation to music practice, that is, in relation to the so-called Common Practice Notation (even if extended, see later) and its performance. 
A filtering process is thus required to perform data reduction. Once the CSV file has been parsed, the Preprocessor operates in three steps:
\begin{enumerate}
    \item the SNR parameter is used again to filter out more data. With SNR$= 125$, $843$ (over $\sim26$k) sources remain available, the most relevant ones.
    \item spatial dimensions are reduced to $2$ by keeping distance and $\phi$ while discarding $\theta$. This results in a 2D projection of the  original space. 
    \item apart from spatial coordinates, frequency and GW amplitude are the sole parameters that are kept, while the other are discarded.
\end{enumerate}
Finally, the Preprocessor normalizes 2D sources position in a $[0.0, 1.0]$ range for sake of simplicity, so that SNR-filtered sources occupy the whole normalized space.
The Preprocessor can be tuned interactively by coding in SuperCollider. 

\subsection{Stage II: Space modeler}
\begin{figure}[t]
\centering
\includegraphics[width=0.9\columnwidth]{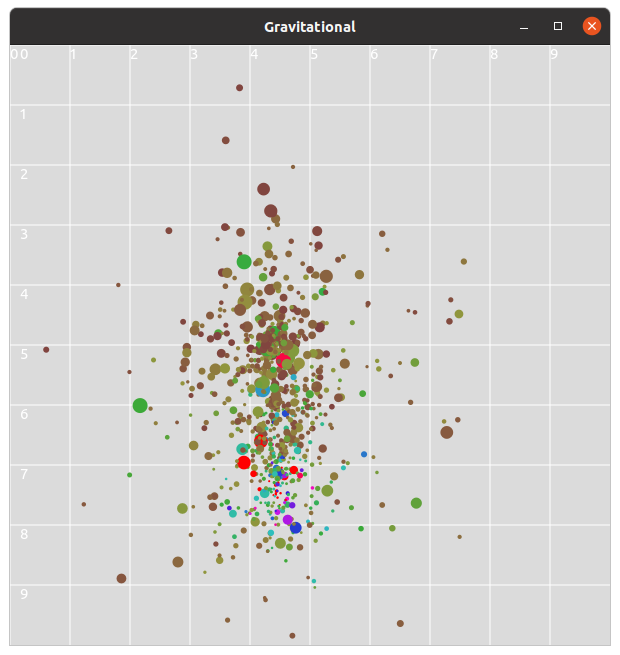}
\caption{2D display for the SNR filtered sources. Each circle represents a sound/gravitational wave source with a radius proportional to the source frequency. The color coding reflects the source amplitude: the amplitude increases from brown/green, to blue/violed to red.\label{fig:gravit}}
\end{figure}

Data reduction performed on source data by the Preprocessor allows to obtain a 2D visual display useful for a compact exploration of the overall data organization. While visual displaying is independent from sonification (which can be thought of as a form of aural displaying), if the two strategies are consistent it becomes easier to explore the data set, in order to understand its features. Figure \ref{fig:gravit} shows the 2D space in the interactive SuperCollider GUI.
Each source is represented by a dot circle with the radius proportional to frequency. The color of each circle is obtained by using a hue-saturation-value color model, in which the three dimensions are scaled proportionally to the source amplitude. This can be read as: the highest is the amplitude, the more saturated and brightest is the color. In terms of hue, amplitude increases from brown/green to blue/violet to red.
Again, interactive control on visualization is made possible by code in the SuperCollider environment.

In this spatial model, the sonification metaphor is based on considering each GW source (i.e.  circle in the GUI) as a sound emitting source.
Each of these sound sources is provided with two features: a characteristic frequency and an emission frequency.
Characteristic frequency indicates a specific pitch while emission frequency is related to a repetition rate. In short, each source is given a ``pitch'' to be repeated at a certain rate. The resulting acoustic space is thus meant as a set of pulsating, pitched sources. In terms of parameter mapping, the GW frequency is mapped onto the emission frequency (rate), while the GW amplitude is inversely mapped onto characteristic frequency (pitch), so that highest amplitudes result into lowest pitches. Such an arrangement directly derives from the observation of the organization of the sources in space. Sparse sources at the periphery of the space will typically have low GW amplitude yielding high pitches. The center of the space is dense and includes most of sources with highest amplitudes. The sonification of this region will consequently feature the lowest notes together with high ones.

\subsection{Sequencer and Explorer}
\begin{figure}[t]
\centering
\includegraphics[width=0.9\columnwidth]{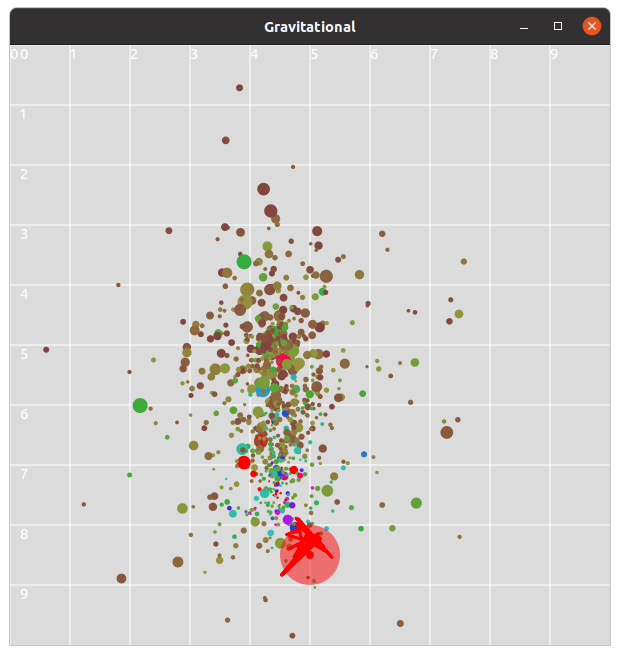}
\caption{Trajectory (red broken line) and audibility radius (red circle).\label{fig:arci}}
\end{figure}

This abstract acoustic space is to be navigated by an Explorer. The main metaphor is travelling through the space and detecting sources while approaching them. The Explorer is defined by a trajectory, to be traversed at constant speed, and by an audibility radius $a_r$. 
Figure \ref{fig:arci} shows a trajectory (red line) with $a_r$ for its starting point (red circle), showing the audible sources for that position.

Only sources inside $a_r$ are taken into account, the other outside it being discarded. The spatial metaphor is perceptually viable: sources near the Explorer are audible, while sources beyond $a_r$ remain silent.
The Explorer is able to detect a source within its audibility radius and measure its distance. 
Such a mechanism has some interesting features on the composition side:
\begin{itemize}
    \item trajectories introduce a time-based behavior in the space;
    \item trajectories set the pace of the music (though the control of their length and speed);
    \item trajectories select which part of and how the space is to be heard, e.g. they can be tuned by observing the visual display;
    \item $a_r$ represent a further filtering over sources;
    \item distance between Explorer and each source provides a viable, ecological parameter to describe sonic transformations.
\end{itemize}
The trajectory/audibility radius framework has been introduced by ~\cite{DBLP:conf/cmmr/ValleLS09} in relation to dynamic soundscape simulation. Such an ecological metaphor~\cite{Bardelli2021} provides a link between source data and auditory perception and places the composition technique, even if indirectly, in the context of ``soundscape composition''~\cite{Westerkamp:2002}. To sum up, GW sources are thought as elements contributing to an abstract but motivated soundscape that changes dynamically as a listener moves along an exploring path.
In the space model, trajectories allow to explore different parts of the space while variably configuring the available parameters, e.g. $a_r$ can be set in relation to each trajectory. Thus, the final piece {\em Periplo del latte} is a cycle made up of shorter sections, each one implementing an exploration along a specific trajectory. Various algorithms have been designed to construct trajectories. As an example, the trajectory in Figure \ref{fig:arci} results from a Brownian motion constrained by a density parameter, so that new points always bounce inside a certain region. This feature makes it possible to explore a limited set of sources in a dynamical manner.\\
With reference to Figure \ref{fig:detailSpace}, the Sequencer is passed the SpaceModel and the trajectory created by the Explorer. The Sequencer generates sequences of events for sources inside $a_r$ while moving along the trajectory over time. For each active event (which rate depends on the source frequency), the attack time is retained and the GW amplitude is mapped logarithmically in the audio frequencies corresponding to the MIDI range $[21, 108]$ (piano pitches). Also, the distance with the actual point on the trajectory is computed, so that the output format for each event is \verb# att, freq, dist#, this information being then logged to the \verb#ev# CSV file.\\
The whole composition process can be fine tuned by visualizing in real time trajectories while simulating the sound output by means of a synthesized piano (GUI and Audio in Figure\ref{fig:detailSpace}).
Apart from sound simulation, and even if the output acoustic frequencies are already constrained in the piano range ($[27.5, 4186] Hz$), the final result of stage III is still abstract from sound implementation, and the resulting sequenced events can be used to feed e.g. a digital sound synthesis process, even in real time. In this case, as the software architecture is modular, it is easy to replace a previously computed trajectory with a point input by the user. 

\section{M(usic)Interpretation}
\begin{figure}[t]
\centering
\includegraphics[width=0.9\columnwidth]{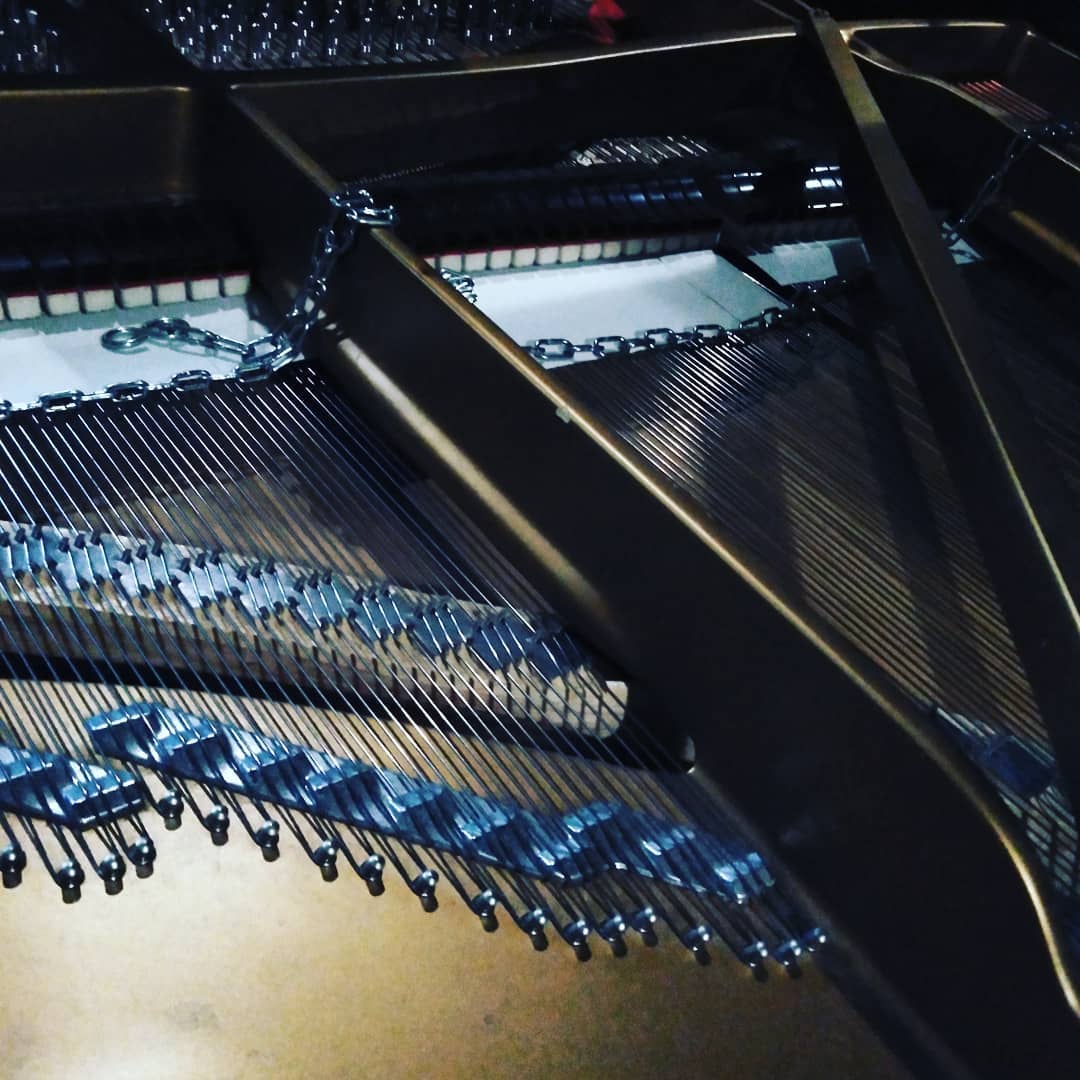}
\caption{Piano preparation.\label{fig:prepPiano}}
\end{figure}

The main project constraint was to write for piano solo. In order to disconnect piano from its standard music associations and to create a mood of otherness related to such distant and literally invisible objects as gravitational wave sources, the piano has been prepared (Figure \ref{fig:prepPiano}). Metal chains have been inserted across the string length for the whole string set, so to produce a bright, buzzing (rather than bouncing) timbre that still does not interfere with pitch discrimination. Chain weights are to be accommodated in relation to string diameter, so to ensure a consistent effect all over the strings and to be consistently sensitive to dynamics. Also, paper strips are to be placed under all the dampers. These two preparations shift the piano sound towards a tuned percussion instrument while retaining its typical keyboard agility.\\
\begin{figure*}[t]
\centering
\includegraphics[width=2\columnwidth]{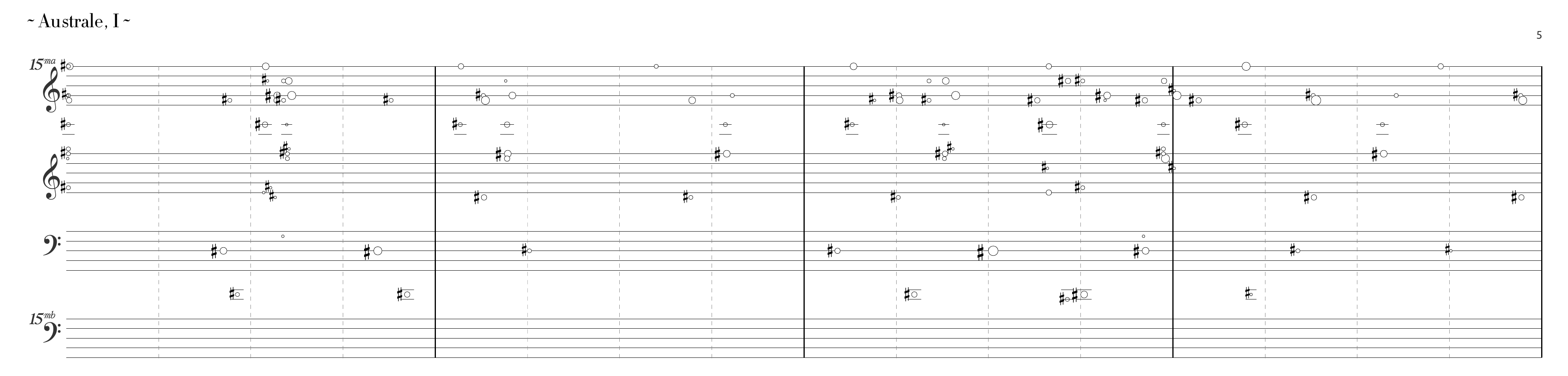}
\caption{{\em Australe, I}, first four measures.\label{fig:australe}}
\end{figure*}
\begin{figure}[t]
\centering
\includegraphics[width=1\columnwidth]{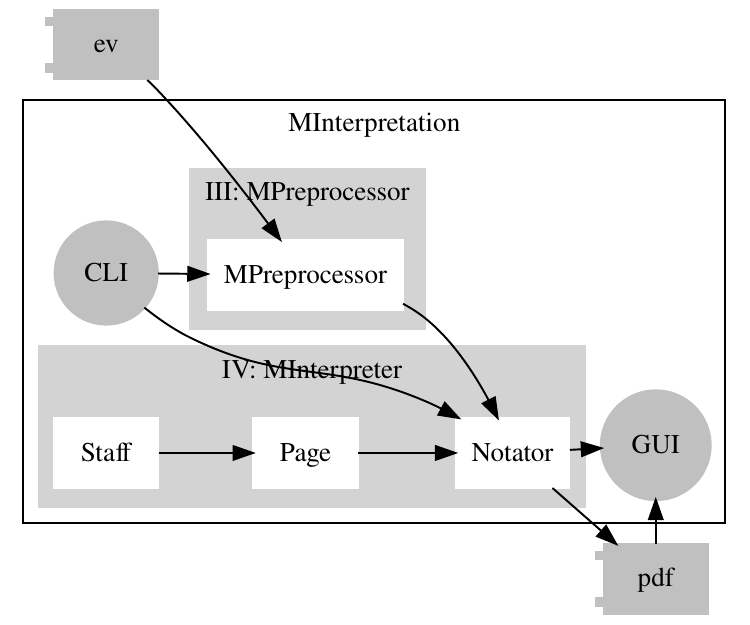}
\caption{Stages III and IV.\label{fig:detailInt}}
\end{figure}
Some other constraints cascade from writing for piano ($pc$), that we will address in the next sections:
\begin{enumerate}
    \item 88 pitches are available in 12-tone equal temperament ($pc_1$);
    \item a theoretical maximum of 10 simultaneous notes are available resulting from key pressed by the player. Moreover, hands positions should be addressed as they constraint the available notes that can be played together ($pc_2$);
    \item a representation format (notation) must be devised that can compactly express the information related to events while still being playable ($pc_3$).
\end{enumerate}
Stages III and IV are entirely implemented in Python\footnote{\url{https://www.python.org/}}, that is --as SuperCollider-- a high level, interpreted language, well fitted for music composition.
\subsection{Stage III: MPreprocessor}
The Sequencer outputs the \verb#ev# log file which is passed to the M(usic)Preprocessor (Figure \ref{fig:detailInt}). For each event, the three associated \verb#att, freq, dist# values are remapped as follows:
\begin{itemize}
    \item the attack of each event is rounded to the nearest value in relation to three available grids quantizing a beat in $5, 6, 8$ subdivisions and assuming a bpm with $\frac{1}{4} = 60$; 
    \item the frequency is converted into its integer MIDI note value, so that it can be mapped onto a single piano key (see $pc_1$);   
    \item distance, which range is comprised in $[0, a_r]$, is mapped linearly on an inverted $5$-step scale representing the event dynamics, so that minimum distance is mapped onto $5$ (= {\em ff}) and maximum one onto $0$ (= {\em pp}). This is indeed a reference, even if largely ``cartoonified''~\cite{Rocchesso:2003}, to intensity as a monaural psycho-acoustic cue for distance detection in space~\cite{Fontana:2002}.
\end{itemize}
\subsection{Notation}
In order to comply with $pc_3$, Common Practice Notation (CPN) has been taken into account as the main reference for the piece. In relation to the former, some issues have emerged. While the music is metrically organized by means of proportions (see before), still parallel irregular grouping are very complex to handle in terms of graphic layout of CPN. Another issue is that large clusters of notes can become cumbersome on a standard piano two-staff system (the so-called ``grand staff''). 
Since the beginning of 20\textsuperscript{th} century, many composers have intensively worked on alternative music notation styles~\cite{Valle:2018}, which are now part of the standard background of contemporary musicians and composers~\cite{Sauer:2009}. Drawing inspiration from piano notation in Helmut Lachenmann's {\em Allegro sostenuto}\footnote{H. Lachenmann, {\em Allegro sostenuto}, Breikopf \& Härtel, Leipzig, 1986/88, KM 2407.}, we thus devised a specific notation format based on four staves, in which two staves surround the standard piano grand staff. Figure \ref{fig:australe} provides an example by reproducing the first four measures of an actual section of {\em Periplo}. Here, the upper and the lower staves extend respectively the treble and the bass clefs, and they sound respectively 2 octaves up and down ({\em 15ma}). Two lines are thus required between the two treble clefs and between the two bass clefs in order to ensure the correct pitch progression. Time representation makes use of the so-called ``time notation'' in which duration is proportional to graphical space in the staff (see ~\cite{Valle:2018}, chap. 3). In Figure \ref{fig:australe}, each measure represents four seconds. In short, chronometric time is represented by assuming  a meter $= \frac{4}{4}$ and a tempo of {\em $\frac{1}{4}$ = 60} bpm. In the notation, dashed vertical lines represent 1-second durations ($= \frac{1}{4}$) and are provided as a reference for the player. In the performance, tempo can be slowed, provided that it is constant over each section. 
Notes are represented as white circles. White color is not related to duration as in CPN. In fact, duration is not specified at all in the notation: it must be accommodated by the player in relation to performance and musical needs (including e.g. the use of the pedal). On the contrary, for each note the attack time is notated precisely, and proportionally to space.  Each note’s attack is represented by its foremost left point on its diameter. As notation is proportional and not metric, the three proportions ($5, 6, 8$) are not explicitly notated, rather they represent a sort of hidden layer of time organization emerging in the performance. For each note, the circle diameter indicates its dynamics, from smallest (= {\em ppp}) to largest (= {\em fff}). In this way, dynamics progression (crescendo/diminuendo) for repeated notes with the same pitch can be easily detected in the score. Alterations are notated exclusively by means of sharp accidentals ($\sharp$). They apply only for the single note for which they are specified. The resulting, pulviscular notation is also meant as an apt reference to stars and constellations. 
\subsection{Stage IV: MInterpreter}
Such a custom notation format prevents the usage of available packages for automatic music notation generation (~\cite{Assayag:1999, Kuuskankare:2006, Taube:1997, agostini:2015}, or the Python-based~\cite{conf/ismir/CuthbertA10}). Rather, a specific implementation has been developed, integrating SuperCollider with Python in an architecture that can be though of as ``fluid'' in relation to Computer-Aided Composition systems~\cite{Valle:2008a}.
As shown in Figure \ref{fig:detailInt}, remapped data from MPreprocessor are now suitable for the Notator module that is responsible for the implementation of the notation format discussed above in a completely automatic fashion. Notator handles the generation of the final graphical elements in PDF format by exploiting Shoebot, a Python library for vector graphics\footnote{\url{http://shoebot.net/}}. For each piece, a score is made up of pages, in turn made up of staves, containing custom music symbols. The last two layers are thus handled by two separate Python classes, Page and Staff (Figure \ref{fig:detailInt}). Staff implements the custom extended grand staff discussed before. Notator examines the event log and first defines the required number of Page instances. For each Page, it creates two Staff instances. For each event, $att$ is then used to define a certain position in the correct Page and Staff. The vertical position on the Staff depends on the pitch while the radius of the note circle symbol is proportional to dynamics. Graphic generation is fine-tuned interactively by monitoring the generated PDFs and subsequently modifying Python settings. \\
Events and their organization over time are not formally defined per se, rather they result from the exploration of the space, thus making very difficult to predict a priori the actual output. The latter is a particularly severe issue in relation to $pc_2$ (piano realization). Interactive monitoring makes possible a trial-and-error approach to assess the overall density of notes in terms of performing possibility, modifying the input parameters if the score cannot be played (e.g. too many notes or too distant finger positions). As a safeguard, in case of complex blocks of notes, the piano player is explicitly left the decision whether to omit some notes for sake of feasibility. 
\\ All the pages resulting from automatic notation generation are then assembled by scripting Con\TeX, a \TeX-based package for document preparation\footnote{\url{https://wiki.contextgarden.net/}}. The final score of {\em Periplo del latte} consists of 11 sections. Figure \ref{fig:australe} shows the first four measures (top half-page in landscape format) of the {\em Australe, I} section. The piece results from the exploration shown in Figure \ref{fig:arci}.

\section{Conclusions}
Our sonification project had to address many different issues: from a large amount of data to their spatial internal organization, from instrumental composition to automatic generation of music notation. The modular four-stage architecture that we have discussed has allowed us to cope with such issues so to integrate them into a unified sonification model. 
{\em Periplo del latte} debuted in the context of {\em Einstein's Sonata} on October 29\textsuperscript{th}, 2021 at Festival della Scienza, Genoa\footnote{\url{http://www.festivalscienza.it/}. A teaser can be found here \url{https://www.youtube.com/watch?v=q_zZu47ky5I}}. The festival is meant to overcome the traditional opposition between scientific and humanistic culture and its intended audience is as general as possible. {\em Einstein's Sonata} has been preceded by a talk discussing both the original experiment and the sonification process. This public display has demonstrated the potential of sonification not only as a scientific data display tools, but in our case as a way to bring a general public closer to advanced scientific concepts by means of music while respecting the scientific nature of the project. At the same time, it has allowed us to introduce sonification per se, largely unknown to a wider audience.

The recording of the piece is available at\\ \url{https://andreavalle.bandcamp.com/album/periplo-del-latte}.

\begin{acknowledgments}
The authors would like to thank the team of the {\em Einstein's Sonata} project, Samantha Stella (artistic director), Luca Ieracitano (pianist), and Nicola Tamanini, who contributed to the project at initial stages.
This works has been funded by the by the Gruber and Rubicon postdoctoral fellowships (grant number 019.183EN.015) awarded to Valeriya Korol respectively by the International Astronomical Union (IAU) and the Netherlands Research Council (NWO).

\end{acknowledgments}

\bibliography{miaBiblio}

\end{document}